\documentclass[aps,pra,showpacs,amssymb,superscriptaddress,twocolumn,longbibliography]{revtex4-2}

\usepackage[T1]{fontenc}
\usepackage[english]{babel}
\usepackage[utf8]{inputenc}
\usepackage{amsmath,amssymb, amsfonts}
\usepackage[colorlinks=true,citecolor=blue]{hyperref}
\usepackage{graphicx}
\usepackage{svg}
\usepackage{amsmath}
\usepackage{slashed}
\usepackage{bbm}
\usepackage{soul}
\usepackage{array,booktabs}
\usepackage{natbib}
\usepackage{bm}
\usepackage{bbold}
\usepackage{hhline}
\usepackage{xcolor}
\usepackage{mathtools} 
\usepackage{float}
\usepackage{tabularx}
\usepackage{tikz}
\usepackage{physics}

\usepackage[capitalize]{cleveref} 

\usepackage{MnSymbol}

\newcommand{\id}{\ensuremath{\mathbbm{1}}}

\newcommand{\osaFig}[1]{}

\begin{document}

\title{Certifying high-dimensional quantum channels}

\author{Sophie Engineer*}
\email[Email address: ]{sophie.engineer@bristol.ac.uk}
\affiliation{Institute of Photonics and Quantum Sciences (IPAQS), Heriot-Watt University, Edinburgh, United Kingdom}
\affiliation{Quantum Engineering Centre for Doctoral Training, H.\ H.\ Wills Physics Laboratory and Department of Electrical \& Electronic Engineering, University of Bristol, Bristol, United Kingdom}

\author{Suraj Goel*}
\affiliation{Institute of Photonics and Quantum Sciences (IPAQS), Heriot-Watt University, Edinburgh, United Kingdom}

\author{Sophie Egelhaaf*}
\affiliation{Department of Applied Physics, University of Geneva, Switzerland}

\author{Will McCutcheon}
\affiliation{Institute of Photonics and Quantum Sciences (IPAQS), Heriot-Watt University, Edinburgh, United Kingdom}

\author{Vatshal Srivastav}
\affiliation{Institute of Photonics and Quantum Sciences (IPAQS), Heriot-Watt University, Edinburgh, United Kingdom}

\author{Saroch Leedumrongwatthanakun}
\thanks{Current address: Division of Physical Science, Faculty of Science, Prince of Songkla University, Hat Yai, Songkhla 90110, Thailand.}
\affiliation{Institute of Photonics and Quantum Sciences (IPAQS), Heriot-Watt University, Edinburgh, United Kingdom}

\author{Sabine Wollmann}
\affiliation{Institute of Photonics and Quantum Sciences (IPAQS), Heriot-Watt University, Edinburgh, United Kingdom}

\author{Benjamin D.M. Jones}
\affiliation{Quantum Engineering Centre for Doctoral Training, H.\ H.\ Wills Physics Laboratory and Department of Electrical \& Electronic Engineering, University of Bristol, Bristol, United Kingdom}

\author{Thomas Cope}
\affiliation{Institut f{\"u}r Theoretische Physik, Leibniz Universit{\"a}t Hannover, 30167 Hannover, Germany}

\author{Nicolas Brunner}
\affiliation{Department of Applied Physics, University of Geneva, Switzerland}

\author{Roope Uola}
\affiliation{Department of Applied Physics, University of Geneva, Switzerland}

\author{Mehul Malik}
\email[Email address: ]{m.malik@hw.ac.uk}
\affiliation{Institute of Photonics and Quantum Sciences (IPAQS), Heriot-Watt University, Edinburgh, United Kingdom}

\begin{abstract}

The use of high-dimensional systems for quantum communication opens interesting perspectives, such as increased information capacity and noise resilience. In this context, it is crucial to certify that a given quantum channel can reliably transmit high-dimensional quantum information. Here we develop efficient methods for the characterization of high-dimensional quantum channels. We first present a notion of dimensionality of quantum channels, and develop efficient certification methods for this quantity. We consider a simple prepare-and-measure setup, and provide witnesses for both a fully and a partially trusted scenario. In turn we apply these methods to a photonic experiment and certify dimensionalities up to 59 for a commercial graded-index multi-mode optical fiber. Moreover, we present extensive numerical simulations of the experiment, providing an accurate noise model for the fiber and exploring the potential of more sophisticated witnesses. Our work demonstrates the efficient characterization of high-dimensional quantum channels, a key ingredient for future quantum communication technologies. 

\end{abstract}

\maketitle

{\it Introduction.---} Quantum communication networks promise to revolutionize information processing, communication and metrology \cite{kimble2008quantum}. Quantum channels represent a key ingredient of these networks, allowing transmission of information between remote users, and their certification represents a key challenge for the development of future quantum communication technologies. 

However, the practical characterization of quantum channels faces a number of challenges. First, real-world quantum channels are usually complex physical systems (e.g.~multi-mode fibers), the full characterization of which is a daunting task~\cite{bouchard2019quantum,popoff2010measuring,carpenter2015observation}. Second, estimating the quantum channel capacity, even for simple and fully characterized channels, is extremely challenging~\cite{nielsen2010quantum}. 

To overcome these challenges, a number of approaches have been developed to certify specific properties of interest of quantum channels. First, a series of works have develop practical tests for obtaining lower bounds on the quantum channel capacity, see for e.g. \cite{macchiavello2016detecting,macchiavello2016witnessing,Pfister2018}. Another approach consists in certifying the quantum nature of a channel (i.e. that it can be used to distribute entanglement) \cite{Pusey2015,Rosset2018,graffitti2020measurement,mao2020experimentally}. More recently, these questions have also been investigated in the device-independent (black box) setting \cite{DallArno2017,Wagner2020,Sekatski2023}. 

So far, most of this research has focused on the simplest qubit channels. Very little is known beyond this case \cite{ringbauer2018certification}, in particular for the certification of high-dimensional (HD) channels, i.e. channels supporting the transmission of high-dimensional quantum systems (qudits)~\cite{valencia2020unscrambling}. Notably, this question is well motivated by recent developments on using HD quantum systems for communication, allowing for boosted information capacity and increased noise/loss resilience, see e.g. \cite{srivastav2022quick,zhu2021high,cozzolino2019high}. In particular, these features of HD systems may enable secure quantum communication in noise and loss regimes where qubit systems would be insecure \cite{Mirhosseini2015qkd}.

In the present work, we address the question of certifying the dimensionality of an HD quantum channel. That is, how can we quantify the ability of an uncharacterized quantum channel to transmit genuinely HD quantum information. First, we present a theoretical framework to discuss these questions. We define a notion of dimensionality of a quantum channel and develop efficient methods for testing this channel dimension in practice, via a prepare-and-measure setup (i.e.~not requiring entanglement). In turn, we demonstrate the experimental relevance of these methods by testing the dimensionality of commercial multi-mode fibers (MMF) for the transmission of HD systems encoded in the transverse-spatial degree-of-freedom of photons. For example, we certify a minimum dimensionality of $59$ using only two measurement bases. Moreover, we present extended numerical simulations of idealised and noisy HD channels to verify our results, and to show that protocols with more bases lead to a significant improvement in certified dimension.

\begin{figure}[t!]
\centering\includegraphics[width=0.45\textwidth]{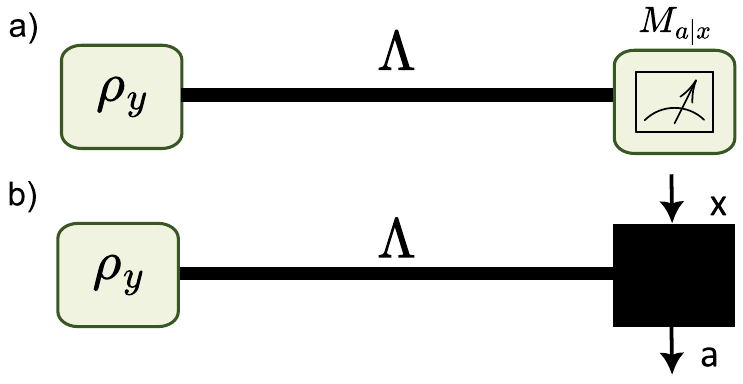}
\caption{We discuss the task of bounding the dimensionality of an unknown high-dimensional quantum channel $\Lambda$. A prepare-and-measure setup is considered, where a sender prepares various input states $\rho_y$. We develop witnesses for two scenarios with varying level of trust on the receiver's device: a) A fully trusted model, where the measurement operators $M_{a|x}$ are characterized, and b) A partially trusted model, where the measurement device is uncharacterized.
}
\label{fig:schematic}
\end{figure}

{\it Theory.---} We consider a prepare, transmit, and measure setup as shown in Fig.~\ref{fig:schematic}. A sender (Alice) prepares an input quantum system in a number of possible quantum states $\rho_y$, where the label $y$ refers to the choice of input state. In turn, $\rho_y$ is transmitted to a receiver (Bob) via a quantum channel $\Lambda$, a completely positive trace-preserving (CPTP) map, resulting in the output state $\rho^{out}_y = \Lambda(\rho_y)$. Bob then performs a measurement represented by a set of POVMs $\{M_{a|x} \}_{a,x}$, where $x$ denotes the choice of measurement and $a$ its outcome. The resulting input-output statistics are then given by 
\begin{align} \label{eq:statistics_probs}
    P(a|x,y) = \operatorname{Tr} (\Lambda(\rho_y) M_{a|x} ) .
\end{align}

Our main interest here is in certifying certain properties of the quantum channel $\Lambda$ based on the observed statistics. In particular, we consider experiments in which the prepared states of Alice and the measurements of Bob involve HD quantum systems (let $d$ denote the system's Hilbert space dimension). We test the ability of the channel to transmit this HD quantum information faithfully. Indeed, in the best possible case, the channel would perfectly transmit the states $\rho_y$ to Bob, in which case $\Lambda$ simply corresponds to a ($d$-dimensional) identity map. Of course, real-world channels are subject to unavoidable imperfections such as noise, loss, and cross-talk, which may significantly affect their ability to transmit quantum information. For example, when the noise is large enough, the channel is entanglement-breaking (in the sense that it can no longer be used to transmit any entanglement) and thus becomes ``classical,'' as it could be replaced by classical communication in a measure and (re)-prepare strategy \cite{horodecki2003entanglement}.

Our focus here is on the intermediate regime, where the channel degrades the HD quantum information (or equivalently HD entanglement) without destroying it completely. For example, maximal $d$-dimensional entanglement may not survive, but lower $k$-dimensional ($1 \leq k \leq d$) entanglement could be transmitted. To quantify this feature, we consider the following figure of merit, the so-called Schmidt number of a quantum channel~\cite{chruscinski2006partially} given by
\begin{align} \label{eq:channel schmidt number}
    \text{SN}(\Lambda)=\min\max_k \text{Rank}(K_k),
\end{align} 
where $\{K_k\}_k$ are Kraus operators and the Schmidt number can be viewed as the dimensionality of the channel $\Lambda$. Importantly, to be meaningful, this quantity must be defined by considering all possible implementations of the channel in terms of its decomposition into Kraus operators $\{K_k\}_k$, such that $\Lambda(\cdot) = \sum_k K_k (\cdot) K_k^\dagger$. Then one should minimise (over decompositions) the largest rank of any of the Kraus operators $K_k$. Loosely speaking, the quantity $\text{SN}(\Lambda)$ captures the dimension of the largest subspace (within the input Hilbert space) that can be coherently transmitted through the channel.

Our goal now is to construct methods for lower bounding the channel Schmidt number based on measurement statistics. For this, we take advantage of the celebrated Choi–Jamiołkowski (CJ) isomorphism, which associates to every quantum channel $\Lambda$ a bipartite quantum state $\rho_\Lambda$. Notably, the Schmidt number of a channel is equal to the Schmidt number of the corresponding state, i.e. $\text{SN}(\Lambda) = \text{SN}(\rho_\Lambda)$. Using recently developed methods for entanglement detection in HD systems \cite{Bavaresco2018,designolle2021genuine,morelli2023witness}, we can construct effective witnesses for lower bounding the channel Schmidt number. 

We consider two different scenarios with varying levels of trust. First, we start with a full trust (FT) model, where the input states $\rho_y$ and the measurement operators $\{ M_{a|x}\}$ are fully characterized. Then we move to a scenario with partial trust (PT), where we require only trust on the input states. In both cases, we construct witnesses for the channel Schmidt number based on mutually unbiased bases (MUBs). Recall that a set of bases in $\mathbb{C}^d$ is called MUB if, for any pair of bases, $| \braket{\phi_i}{\psi_j}| = 1/ \sqrt{d}$ for any vector $\phi_i$ within the first basis and any vector $\psi_j$ within the second basis. When $d$ is a prime power, there exist sets of $d+1$ MUBs \cite{Wootters1989}. To define the witnesses, it is useful to denote the prepared state via a double index, i.e. $\rho_{b|y}$, where $y$ indicates the chosen MUB and $b$ denotes the eigenvector. Specifically, we are then interested in the terms
\begin{align} \label{eq:statistics}
    C_{a,b|x} = \operatorname{Tr} (\Lambda(\rho_{b|x}) M_{a|x} ) ,
\end{align}
giving the correlations in the $x^\text{th}$ MUB. Note that here, all states and measurements correspond to MUBs.

\begin{figure*}[t!]
    \centering
\includegraphics[width=\textwidth]{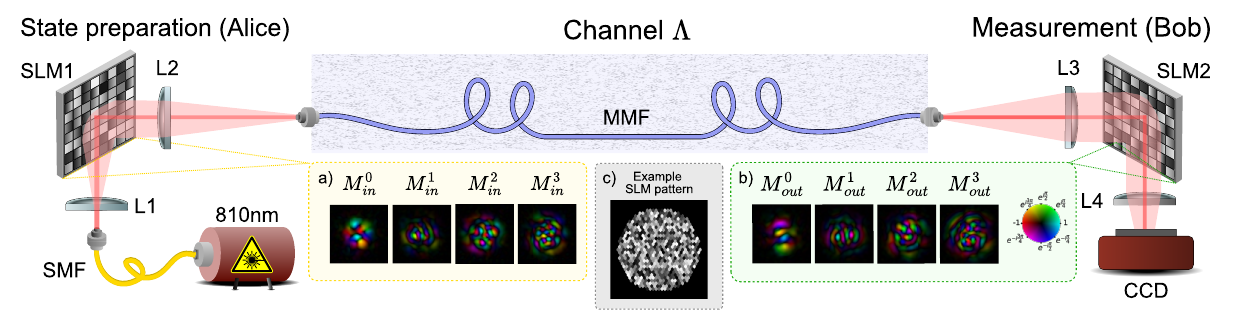}
\caption{Experimental setup. The experiment consists of three parts: state generation, state propagation through the channel and state measurement. To generate the input state, a spatially-coherent light
source (810 nm) is incident on a programmable phase-only spatial light
modulator (SLM1). This is then coupled into a graded-index (GRIN) multi-mode fiber (MMF), representing a noisy channel. The state at the output of the MMF is incident on SLM2 followed by a CCD camera, the combination of which allows one to perform projective measurements on the output state. The modes generated correspond to the eigenmode basis of the fiber and a MUB with respect to the eigenmode basis. (a) and (b) insets display four examples of input and output modes, arranged in descending order of eigenvalue magnitude. (c) shows an example of a phase pattern displayed on the SLM's.  }
\label{fig:experiment}
\end{figure*}

Let us start with the FT scenario. In the entanglement picture (i.e.~via the CJ isomorphism) this corresponds to the usual ``trusted device'' scenario for entanglement detection. We can therefore adapt the SN witness from Ref. \cite{Bavaresco2018} to obtain the following inequality:
\begin{align} \label{eq:Mubbywitness}
    &\sum_{a=0}^{d-1} (C_{a,a|0} + d C_{a,a|1}) - \sum_{\substack{a,a',b,b'=0 \\ a\neq a',a\neq b,\\b\neq b',b'\neq a'}} ^{d-1} \gamma_{a,a'}^{b,b'}\sqrt{C_{a',b'|0}C_{a,b|0}} \leq d(n+1) 
\end{align}
with $\gamma_{a,a'}^{b,b'}=1$ if $(a-a'-b+b')\text{(mod } d)=0$ and $\gamma_{a,a'}^{b,b'}=0$ otherwise (see Appendix~\ref{SM:FTwitness_derivation} for derivation). Importantly, this inequality must hold for any channel with $\text{SN}(\Lambda)\leq n$. Hence a violation of the inequality implies that the channel $ \Lambda$ must have a strictly larger Schmidt number, i.e. $\text{SN}(\Lambda)> n$.

Now let us move to the PT scenario. Here we wish to characterize the quantum channel while relaxing the trust on Bob's measurement device. That is, we consider the latter to be error prone; however we do not consider the device to be controlled by a malicious adversary as in the device-independent scenario. Hence we now relax the assumption that the measurement operators correspond to MUBs, and consider Bob's measurement device as uncharacterized. Going to the entanglement picture, this corresponds to a scenario for entanglement detection in the one-sided trusted model, where the measurement device is trusted for one party but untrusted for the other. This corresponds to the task of quantum steering, for which Schmidt number witnesses have recently been developed \cite{designolle2021genuine,Designolle2022,deGois2023}. From Ref.~\cite{designolle2021genuine}, we obtain the witness 
\begin{align} \label{eq:channel schmidt number bound}
    \sum_{x=1}^2 \sum_{a=0}^{d-1} C_{a,a|x}  \leq \frac{2 \sqrt{n} (d + \sqrt{d})}{\sqrt{n} + 1},
\end{align}
for the case where a pair of MUBs is tested. Again, observing a violation of the above inequality implies that the underlying channel $\Lambda$ must have $\text{SN}(\Lambda)> n$ \footnote{More precisely, a violation of \eqref{eq:channel schmidt number bound} implies that the effective measurements $\{ \Lambda^*({M_{a|x}})\}$ are strongly incompatible (in the sense of being non $n$-simulable \cite{ioannou2022simulability}), which implies that both the channel $\Lambda$ and the measurements $\{M_{a|x}\}$ are high-dimensional.}.

{\it Certification of multi-mode fiber dimensionality.---} In our experiment the quantum channels to be characterized are two commercial graded-index MMF optical fibers with lengths of 2~m and 5~m. Fig.~\ref{fig:experiment} shows a schematic of the experimental setup. The input states prepared by Alice are encoded in the transverse-spatial macro-pixel basis. The desired input MUB states are created via a spatial light modulator~(SLM) and lens system. The MUB measurements of Bob are implemented using an optical system consisting of an SLM, lens system, and camera. The measurement apparatus is used to perform both a projective measurement $M_{a|x}$ via computer-generated holograms, as well as a multi-outcome measurement for estimating the transmission matrix (TM) of the fiber. For experimental details, see Appendix~\ref{SM:exp_details}.

Due to modal dispersion, each MMF acts as a mixing channel $\Lambda_{\text{MMF}}$. Our goal is to certify a lower bound on their Schmidt number $\text{SN}(\Lambda_{\text{MMF}})$. We consider both the FT and PT scenarios, testing the witnesses in \eqref{eq:Mubbywitness} and \eqref{eq:channel schmidt number bound}, respectively. 

For each MMF, we first identify the basis that diagonalizes the channel, i.e.~the SVD-basis. To do so, we estimate the TM of the channel using a multi-plane neural network (MPNN) that is trained by the state and measurements on randomised bases~\cite{goel2023referenceless} (see Appendix~\ref{SM:exp_procedure} for details). In turn, we perform a singular-value decomposition on the estimated TM, to obtain the SVD-basis consisting of approximately 200 modes with non-vanishing transmission through the MMF (in a single polarization channel). Note that this basis includes the optical system coupling into and out of the MMF as well as its non-ideal nature, i.e.~bending and experimental imperfections. We stress that knowledge of the TM is not a necessity for testing the witnesses, but it leads to better bounds on the certified dimensionality. As a matter of fact, our experiment can only be approximately described by a single dominant monochromatic TM due to the non-zero spectral bandwidth of our laser source, the finite temporal response of our camera, and the dispersion in the tested MMF.  

\begin{figure*}[t!]
    \centering
\includegraphics[width=\textwidth]{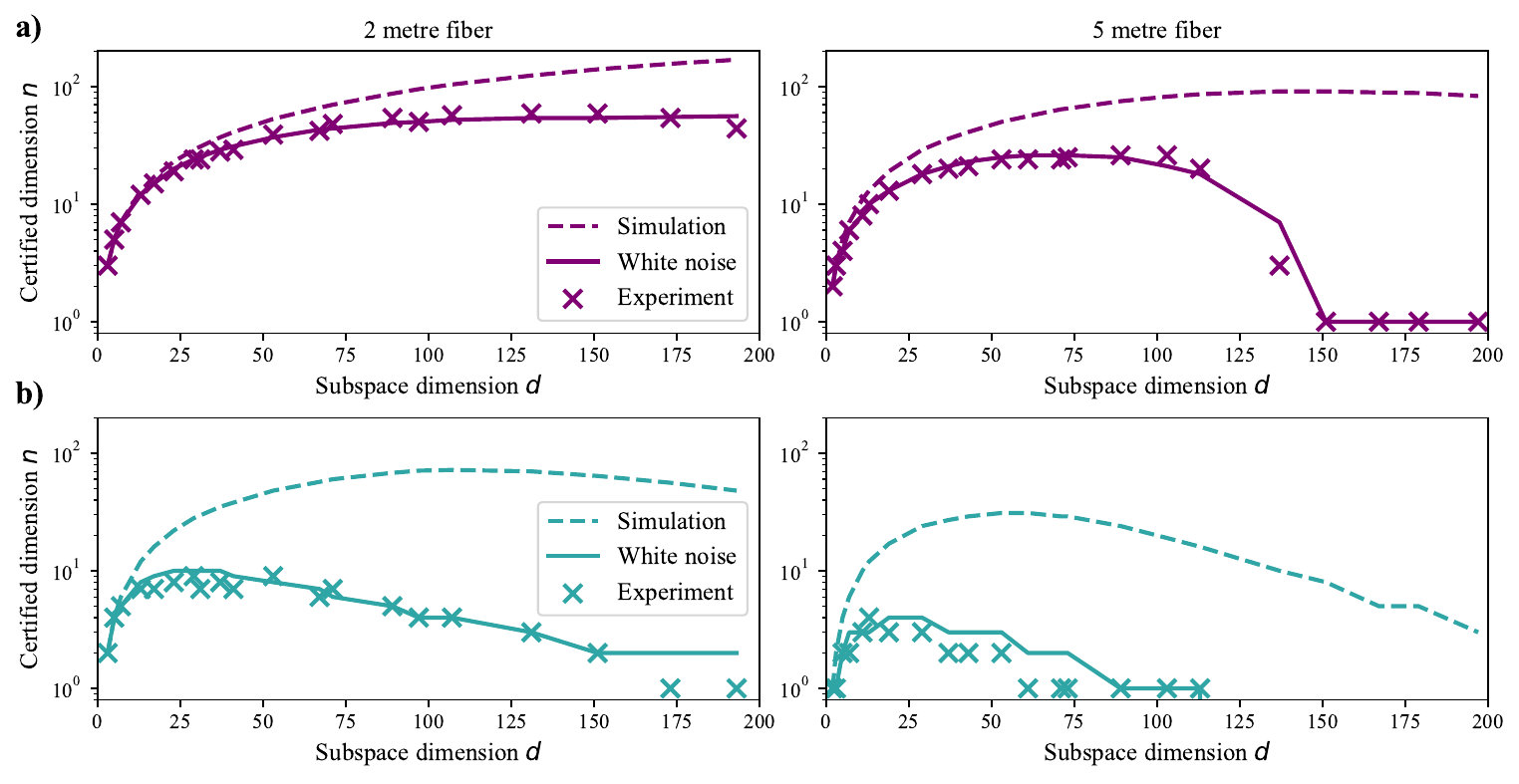}
\caption{Certified dimension $n$ of 2~m and 5~m-long graded-index multi-mode fibers, using the FT witness (Eq.~\ref{eq:Mubbywitness}) and the PT witness (Eq.~\ref{eq:channel schmidt number bound}) when utilising subspaces of different dimension $d$. In a) we show the results of the FT witness (purple) and in b) the PT witness (blue). For all cases, we present the certified dimension for three scenarios: simulation of an idealised fiber accounting for dispersion effects in the MMF (dashed lines), simulation of an idealised fiber with additional white noise (solid lines), and the experimental results (crosses). 
}\label{fig:results}
\end{figure*}

For each MMF, we collect data using two measurement bases: the SVD-basis described above, and a second mutually unbiased basis. This allows us to evaluate the quantities $C_{a,b|x}$ necessary for testing both witnesses, focusing on the case of two MUBs. Testing witnesses on the entire $\approx 200$ modes supported in the fiber can suffer from decreasing coherence between lower- and higher-order modes, whilst smaller subspaces, of a few lower-order modes with similar propagation constants for instance, can better maintain coherence. It is therefore important to perform the analysis considering different subspace dimensions $d$ for the input states and measurements by selecting only the leading modes of the SVD-basis. We stress that this is possible as we assume the input state to be trusted, thus making the considered subspace dimension $d$ a controllable parameter.

The experimental results are presented in Fig.~\ref{fig:results} (crosses), showing the certified dimension $n$, i.e.~a lower bound on the channel Schmidt number $\text{SN}(\Lambda_{\text{MMF}})$, as a function of the subspace dimension $d$. The largest certified dimensions are summarized in Table~\ref{tab:exp_results}. Due to increased dispersion in longer fibers, the certified dimensions are lower for the 5~m MMF. Also, the PT scenario leads to lower certified dimensions compared to the FT as expected, since trust in the measurement device is now relaxed. In all cases, except the FT witness applied to the 2~m fiber, we see a drop in certified dimension $n$ after a critical subspace dimension $d$. This highlights the aforementioned trade-off between increased subspace dimension and including noisy higher order modes.

\begin{table}[b!]
    \begin{center}
    \begin{tabular}{ |c|c|c| } 
     \hline
     Fiber length & Witness & Max. certified dimension ($n$)  \\ 
     \hline
    2~m & FT & 59 (in $d=131$ subspace) \\
    & PT & 9 (in $d=29$ subspace)  \\ 
    \hline
    5~m  & FT &  26 (in $d=89$ subspace) \\ 
    & PT & 4 (in $d=13$ subspace) \\ 
     \hline
    \end{tabular}
    \caption{Maximum experimentally certified dimensions $n$ for each fiber (2~m and 5~m) and witness used (FT and PT).}
    \label{tab:exp_results}
    \end{center}
\end{table}

{\it Simulations with noise and multiple MUBs.---} To gain more insight into our experimental results, we perform simulations of a MMF devoid of manufacturing imperfections or curvatures, but capturing the modal dispersion responsible for channel impurity \cite{shemirani2009principal}. In practice, it is impossible to have channels that perfectly preserve the purity of quantum states. This is due to the spectral bandwidth of the input source, the temporal response of the detectors, and the modal dispersion induced in MMFs. Nevertheless, we study this case to provide an upper bound on the certifiable channel dimensionality one might expect from such idealised fibers. 

When finding the approximate SVD-basis of the fiber from the measured TM (as described in Appendix~\ref{SM:exp_procedure}), we effectively trace over the spectral degree of freedom, leading to a sub-optimal SVD-basis. To simulate this process, we construct a multi-spectral transmission matrix (MSTM) for an idealised fiber with a bandwidth corresponding to that of the laser source in the experiment, i.e.~$810 \pm 1.5$~nm as outlined in Appendix~\ref{SM:MSTM}. We then simulate our experimental estimation of an approximate TM to find the SVD-basis, averaging over the laser bandwidth (see Appendix~\ref{SM:TM_sim} for full details).

From this simulation we generate measurement statistics $C_{a,b|x}$ for each wavelength and then average over the spectral degree of freedom, weighting each term according to the Gaussian envelope of the input laser. This results in an average correlation matrix, which is normalised. Finally, we can estimate the certified dimensions via the FT and PT witnesses. The idealised fiber simulation results are presented in Fig.~\ref{fig:results} (dashed lines). We see that for low subspace dimensions, the experimentally certified dimensions are close to the case of an idealised fiber. For larger subspace dimensions, there is a notable deviation.

This deviation is due to experimental factors such as mode-dependent loss, misalignment, error in state preparation, imperfect measurements and transmission matrix estimation. These effects induce noise in the experiment that increases as the dimension of the considered subspace increases. To model this effect, we consider a simple noise model, where the idealised MMF $\Lambda_{\text{ideal}}$ is (probabilistically) mixed with a completely depolarising channel $\Lambda_{\text{mm}}$ that maps every input state $\rho$ to a maximally mixed output state, i.e. $\Lambda_{\text{mm}}(\rho) = \frac{\mathbb{1}}{d}$. The resulting channel is given by
\begin{equation} \label{eq:noisychannel}
    \Lambda_{\text{noisy}} = p \Lambda_{\text{ideal}} + (1-p) \Lambda_{\text{mm}},
\end{equation}
where $p \in [0,1]$ is the mixing parameter;
for details, see Appendix~\ref{SM:sim_noisy}.

We find good agreement with the experimental data when setting the mixing parameter $p$ to decrease quadratically with respect to the subspace dimension (see Appendix~\ref{SM:sim_noisy} for details). The results of the noise simulation are shown as solid lines in Fig.~\ref{fig:results}. This indicates that a simple noise model can be used and in fact, provides a useful tool for quantifying the noise levels in this type of experiment.

Next, we consider the question of testing Schmidt number witnesses with more MUBs. Indeed, in our experiment we focused on the simplest case of a pair of MUBs. Intuitively, it is clear that a larger dimension could be certified by testing more MUBs. In principle, this is experimentally possible for low subspace dimensions $d$, but quickly becomes challenging when $d$ increases. 

\begin{figure}[t!]
\includegraphics[clip, trim=0.5cm 1.6cm 1.5cm 2.5cm,width=0.47\textwidth]
{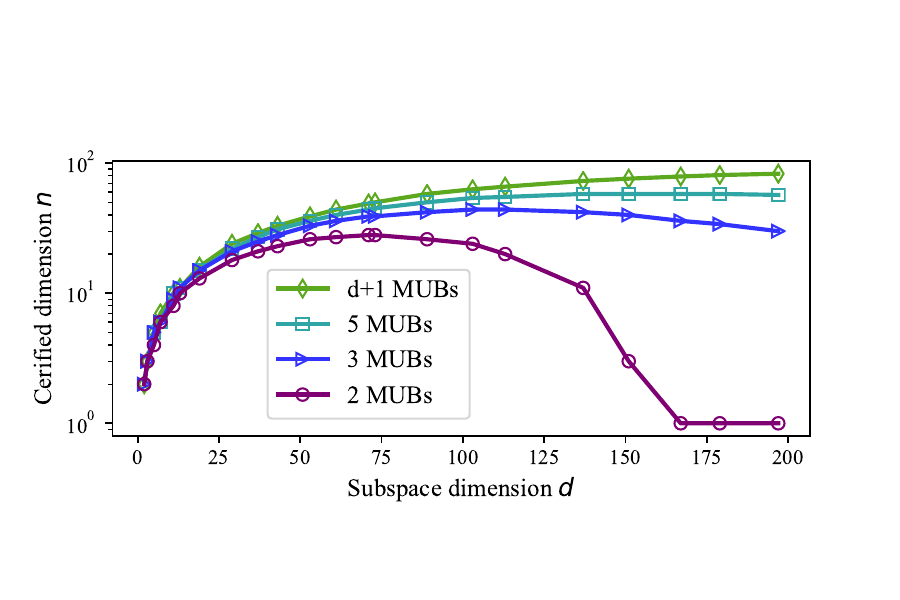} 
\caption{Certified dimension for the simulated idealised 5~m MMF with additional white noise, using the witness of Eq.~\ref{eq:FTwitness} for the fully trusted scenario. The plot shows simulation results as scatter points, with lines added to guide the eye. Notably, considering additional MUBs provides a significant enhancement in the certified dimension.} \label{fig:moreThan2}
\end{figure}

Nevertheless it is relevant to investigate the performance of such more complex witnesses using our simulation model of a noisy MMF. We adapt the witness of Ref. \cite{morelli2023witness} to obtain
\begin{equation} \label{eq:FTwitness}
    \sum_{x=1}^m\sum_{a=0}^{d-1} C_{a,a|x} \leq d+(m-1)n,
\end{equation}
where $n$ is the Schmidt number of the channel and $2\leq m \leq d+1$ denotes the number of MUBs being used, here we choose $m=\{2,3,5,d+1\}$ (see Appendix~\ref{SM:FTwitness_derivation} for derivation). To model the MMF, we use our simulation model for the 5~m fiber (with quadratic noise). The results are presented in Fig.~\ref{fig:moreThan2}, showing a significant increase of the certified dimension. Interestingly, we note also that this advantage increases as the dimension of the subspace grows. This is due to the increase in noise acting on the state. Furthermore, we observe that when testing a fixed number of MUBs (e.g. 3 MUBs), the certifiable dimension starts to drop off when the dimension becomes large. Increasing the number of MUBs, e.g. from 2 MUBs to 3 MUBs, increases the subspace dimension in which the drop off begins and when considering the full set of $d+1$ MUBs, it appears the actual Schmidt number can be certified.

{\it Discussion.---} In this work we developed methods for certifying lower bounds on the dimensionality of quantum channels, as given by the channel Schmidt number. Effective witnesses were derived for this task, tailored to a simple prepare-and-measure scenario with full and partial trust in the output measurements, with the latter being of interest when one does not wish to rely on the output measurements functioning perfectly. We implemented these witnesses in a photonic experiment using commercial MMFs as channels, and reporting certified dimensions up to $n=59$. To get further insight into the problem, we performed extensive numerical simulations of the experiment, providing a simple noise model for the fiber and exploring the potential of more sophisticated witnesses. 

An interesting open question for future work is whether our certification methods could be adapted to characterize other quantities of interest effectively, in particular the quantum channel capacity or the entanglement cost of channels \cite{Berta2013}. Bounding the dimensionality of a quantum channel is directly related to classical concepts such as the channel bandwidth---the maximum data rate a channel can support---which is also limited by modal dispersion. The use of our channel witnesses to bound such classical measures could serve as an interesting topic to be explored in the future. As quantum technologies mature and new quantum networks are established, the need for methods to efficiently characterise quantum channels becomes imperative. Our work takes a significant step in this direction by demonstrating the efficient characterization of high-dimensional quantum channels, a key ingredient in future quantum communication technologies.

{\it Acknowledgements.---} We acknowledge funding from the Swiss National Science Foundation (Ambizione PZ00P2-202179), the Swiss Secretariat for Education, Research and Innovation (SERI) under contract number UeM019-3,  the European Commission under Marie Sklodowska-Curie grant 89224, EPSRC via grant EP/SO23607/1, European Research Council (ERC) Starting Grant PIQUaNT (950402), and the Royal Academy of Engineering Chair in Emerging Technologies programme (CiET-2223-112).

\bibliography{references.bib}

\newpage
\appendix

\begin{widetext}
\section{Derivation of FT witness} \label{SM:FTwitness_derivation}

Recently, witnesses for certifying HD entanglement have been proposed \cite{Bavaresco2018,morelli2023witness}. In the following we show how to adapt such witnesses for constructing witnesses for bounding the Schmidt number of a HD channel. Since here the entanglement witness is based on the assumption of fully characterized measurement operators, we will obtain channel witnesses in the FT scenario.

Let us start by reviewing witnesses for HD entangled states, which provide lower bounds on the Schmidt number of the considered state.
The Schmidt rank (SR) of a pure bipartite quantum state $\ket{\psi}$ is defined as the minimum number of terms needed to express $\ket{\psi}$ as a linear combination of product states. Generalising to mixed states, the Schmidt number of a state $\rho$ is defined as
\begin{align}
    \text{SN}(\rho) := \underset{p_k, ~\ket{\psi_k} }{\min} \max_k \quad&\text{SR}(\ket{\psi_k}) \\\nonumber
    \quad \text{s.t} \quad &\rho = \sum_k p_k \ketbra{\psi_k}{\psi_k} \: .
\end{align}

In order to adapt an entanglement witness to a criterion applicable to channels, we use the Choi-Jamiliokowski (CJ) isomorphism. More precisely, we can then connect the Schmidt number of a channel $\Lambda$ to the Schmidt number of the corresponding ``Choi state'' $\rho_\Lambda$.

We first consider the entanglement witness of Ref. \cite{Bavaresco2018}, which lower bounds the Schmidt number $n$ of a state $\rho$ using the following inequality
\begin{align}
    \sum_{a=0}^{d-1} & \lambda_a^{2} \bra{e_{a|0}e_{a|0}}\rho\ket{e_{a|0}e_{a|0}}
    + \frac{\left(\sum_{a=0}^{d-1}\lambda_a\right)^2}{d}\sum_{a=0}^{d-1} \bra{e_{a|1}e_{a|1}^*}\rho\ket{e_{a|1}e_{a|1}^*} 
    - \sum_{a,b=0}^{d-1} \lambda_a\lambda_b \bra{e_{a|0}e_{b|0}}\rho\ket{e_{a|0}e_{b|0}}& \nonumber \\
    &- \sum_{\substack{a,a',b,b'=0 \\ a\neq a',a\neq b,\\b\neq b',b'\neq a'}}^{d-1} \gamma_{a,a'}^{b,b'} \left(\sqrt{\lambda_a\lambda_{a'}\lambda_b\lambda_{b'}} \sqrt{\bra{e_{a'|0}e_{b'|0}}\rho\ket{e_{a'|0}e_{b'|0}}\bra{e_{a|0}e_{b|0}}\rho\ket{e_{a|0}e_{b|0}}}\right)
    &\leq \sum_{a=0}^{n-1} \lambda_a^2 \\
    &\text{with } \gamma_{a,a'}^{b,b'} = \begin{cases} 1 & \text{ if } (a-a'-b+b')\text{(mod } d)=0 \\ 0 & \text{otherwise}\end{cases}
\end{align}
where $d$ is the dimension, $\{\lambda_a\}_a$ are the Schmidt coefficients of the target state, $\left\{\ket{e_{a|0}}\right\}_a$ is the computational basis and $\left\{\ket{e_{a|1}}\right\}_a$ the MUB basis~\cite{Bavaresco2018}.
In this work the target state is the maximally entangled state $\ket{\psi^+_d}=\frac{1}{\sqrt{d}}\sum_{i=0}^{d-1}\ket{ii}$, i.e. $\lambda_a = \frac{1}{\sqrt{d}}\:\forall a $. This is because for an perfect transmission channel (i.e. the identity channel) the corresponding Choi state is a maximally entangled one. In this case, the above expression can be simplified to 
\begin{align}
    &\frac{1}{d} \sum_{a=0}^{d-1} \bra{e_{a|0}e_{a|0}}\rho\ket{e_{a|0}e_{a|0}} 
    + \sum_{a=0}^{d-1} \bra{e_{a|1}e_{a|1}^*}\rho\ket{e_{a|1}e_{a|1}^*} 
    - \frac{1}{d}
    - \frac{1}{d}\sum_{\substack{a,a',b,b'=0 \\ a\neq a',a\neq b,\\b\neq b',b'\neq a'}}^{d-1} \gamma_{a,a'}^{b,b'} \sqrt{\bra{e_{a'|0}e_{b'|0}}\rho\ket{e_{a'|0}e_{b'|0}}\bra{e_{a|0}e_{b|0}}\rho\ket{e_{a|0}e_{b|0}}} 
    \leq \frac{n}{d} \\
    \Leftrightarrow & \sum_{a=0}^{d-1} \Tr(\rho\ket{e_{a|0}e_{a|0}}\bra{e_{a|0}e_{a|0}})
    + d \sum_{a=0}^{d-1} \Tr(\rho\ket{e_{a|1}e_{a|1}^*}\bra{e_{a|1}e_{a|1}^*})  \nonumber \\
    &- \sum_{\substack{a,a',b,b'=0 \\ a\neq a',a\neq b,\\b\neq b',b'\neq a'}}^{d-1} \gamma_{a,a',b,b'} \sqrt{\Tr(\rho\ket{e_{a'|0}e_{b'|0}}\bra{e_{a'|0}e_{b'|0}}) \Tr(\rho\ket{e_{a|0}e_{b|0}}\bra{e_{a|0}e_{b|0}})} 
    \leq n+1 \quad .
\end{align}
In order to determine the Schmidt number of the channel $\Lambda$ the state $\rho$ is chosen to be the Choi state of the channel, i.e.
\begin{equation}
    \rho_{\Lambda} = (\Lambda \otimes \text{id})(\ket{\psi^+_d}\bra{\psi^+_d}) \quad .
\end{equation}
In general 
\begin{align} \label{eq:substituting Choi state}
    \Tr(\rho_{\Lambda}\ket{e_{a|x}e_{b|x}}\bra{e_{a|x}e_{b|x}})
    &= \Tr((\Lambda \otimes \text{id})(\ket{\psi^+_d}\bra{\psi^+_d}) \ket{e_{a|x}e_{b|x}}\bra{e_{a|x}e_{b|x}}) \nonumber\\
    &= \Tr(\ket{\psi^+_d}\bra{\psi^+_d} (\Lambda^* \otimes \text{id})\left(\ket{e_{a|x}e_{b|x}}\bra{e_{a|x}e_{b|x}}\right)) \nonumber \\
    &= \frac{1}{d} \Tr\left( \Lambda(\ket{e_{b|x}}\bra{e_{b|x}}^T)\ket{e_{a|x}}\bra{e_{a|x}}\right) \: .
\end{align}
Thus, the inequality can be rewritten as
\begin{align}
    \frac{1}{d}&\sum_{a=0}^{d-1} \Tr\left( \Lambda(\ket{e_{a|0}}\bra{e_{a|0}}^T)\ket{e_{a|0}}\bra{e_{a|0}}\right)
    + \sum_{a=0}^{d-1} \Tr\left( \Lambda(\ket{e_{a|1}^*}\bra{e_{a|1}^*}^T)\ket{e_{a|1}}\bra{e_{a|1}}\right) \nonumber \\
    &- \frac{1}{d}\sum_{\substack{a,a',b,b'=0 \\ a\neq a',a\neq b,\\b\neq b',b'\neq a'}}^{d-1} \gamma_{a,a'}^{b,b'} \sqrt{\Tr\left( \Lambda(\ket{e_{b'|0}}\bra{e_{b'|0}}^T)\ket{e_{a'|0}}\bra{e_{a'|0}}\right) \Tr\left( \Lambda(\ket{e_{b|0}}\bra{e_{b|0}}^T)\ket{e_{a|0}}\bra{e_{a|0}}\right)}
    \leq  n+1 \quad .
\end{align}
Due to $\{\ket{e_{a|0}}\}_a$ being the computational basis, $\ket{e_{a|0}}\bra{e_{a|0}}^T = \ket{e_{a|0}}\bra{e_{a|0}}$. Further note that $\ket{e_{a|x}^*}\bra{e_{a|x}^*} = \ket{e_{a|x}}\bra{e_{a|x}}^T$. Hence, the inequality can be simplified in the following way
\begin{align}
    &\sum_{a=0}^{d-1}  \Tr\left( \Lambda(\ket{e_{a|0}}\bra{e_{a|0}})\ket{e_{a|0}}\bra{e_{a|0}}\right)
    + d \sum_{a=0}^{d-1} \Tr\left( \Lambda(\ket{e_{a|1}}\bra{e_{a|1}})\ket{e_{a|1}}\bra{e_{a|1}}\right) \nonumber \\
    &-\sum_{\substack{a,a',b,b'=0 \\ a\neq a',a\neq b,\\b\neq b',b'\neq a'}}^{d-1} \gamma_{a,a'}^{b,b'}\sqrt{\Tr\left( \Lambda(\ket{e_{b'|0}}\bra{e_{b'|0}})\ket{e_{a'|0}}\bra{e_{a'|0}}\right) \Tr\left( \Lambda(\ket{e_{b|0}}\bra{e_{b|0}})\ket{e_{a|0}}\bra{e_{a|0}}\right)} 
    \leq d(n+1) \\
     \Leftrightarrow & \sum_{a=0}^{d-1} \Tr\left( \Lambda(\rho_{a|0})M_{a|0}\right) 
    + d \sum_{a=0}^{d-1} \Tr\left( \Lambda(\rho_{a|1})M_{a|1}\right) 
    -\sum_{\substack{a,a',b,b'=0 \\ a\neq a',a\neq b,\\b\neq b',b'\neq a'}}^{d-1} \gamma_{a,a'}^{b,b'} \sqrt{\Tr\left( \Lambda(\rho_{b'|0})M_{a'|0}\right) \Tr\left( \Lambda(\rho_{b|0})M_{a|0}\right)} 
    \leq d(n+1)
\end{align}
By substituting the definition of $C_{a,b|x}$, we recover the channel witness in Eq.~\ref{eq:Mubbywitness}, i.e.
\begin{align}
     \sum_{a=0}^{d-1} C_{a,a|0}
    + d \sum_{a=0}^{d-1} C_{a,a|1} 
    -\sum_{\substack{a,a',b,b'=0 \\ a\neq a',a\neq b,\\b\neq b',b'\neq a'}}^{d-1} \gamma_{a,a'}^{b,b'} \sqrt{C_{a',b'|0}C_{a,b|0}} 
    \leq  d(n+1) \quad .
\end{align}

A similar procedure can be applied to the witness proposed in~\cite{morelli2023witness} to obtain a witness for channels, given in Eq.~\ref{eq:FTwitness}.
According to Result 1 in~\cite{morelli2023witness} the Schmidt number $n$ of a state $\rho$ is lower bound by
\begin{equation} \label{eq:Morelli}
    \sum_{x=1}^m\sum_{a=0}^{d-1} \Tr\left(\rho \ket{e_{a|x}e_{a|x}^*}\bra{e_{a|x}e_{a|x}^*}\right) \leq 1+\frac{(m-1)n}{d}
\end{equation}
where $\left\{\ket{e_{a|x}}\right\}_a$ are MUBs, $2 \leq m \leq d+1$ is the number of bases used and $d$ is the dimension.
Choosing $\rho = (\Lambda \otimes \text{id})(\ket{\psi^+_d}\bra{\psi^+_d})$ and using~\ref{eq:substituting Choi state} and $\ket{e_{a|x}^*}\bra{e_{a|x}^*} = \ket{e_{a|x}}\bra{e_{a|x}}^T$ yields
\begin{align}
    & \frac{1}{d} \sum_{x=1}^m\sum_{a=0}^{d-1} \Tr(\Lambda(\ket{e_{a|x}}\bra{e_{a|x}})\ket{e_{a|x}}\bra{e_{a|x}}) \leq 1+\frac{(m-1)n}{d} \\
    \Leftrightarrow & \sum_{x=1}^m\sum_{a=0}^{d-1} \Tr(\Lambda(\rho_{a|x})M_{a|x}) \leq d+(m-1)n \\
    \Leftrightarrow & \sum_{x=1}^m\sum_{a=0}^{d-1} C_{a,a|x} \leq d+(m-1)n
\end{align}
which yields the channel witness of Eq.~\ref{eq:FTwitness}.

\section{Derivation of PT witness} \label{SM:steering_derivation}

Following the same ideas as above, we can now present the derivation of our channel witness for the PT scenario, where the measurement operators are now uncharcterized. To do so, we start again from Schmidt number witnesses for entangled states. We now consider a steering scenario, where the measurement operators of one party are characterized, but uncharacterized for the other party. Specifically, we consider the approach in Ref. \cite{designolle2021genuine}, where a criterion for certifying genuine HD steering, providing a lower bound on the Schmidt number of the entangled state. From there, we construct a PT witness for channels. The connection relies again on the equivalence of the Schmidt number of a channel and the Schmidt number of the corresponding Choi state.

Let us assume that the Schmidt number of the considered channel $\Lambda$ is $\text{SN}(\Lambda) =n$.
In the Heisenberg picture of the channel~\cite{kiukas2017}
\begin{equation}
    \Lambda^*(M_{a|x}) = d \Tr_A((M_{a|x}\otimes \id)\rho_\Lambda)^T
\end{equation}
where the Choi state $\rho_\Lambda$ has Schmidt number $\text{SN}(\rho_\Lambda)\leq n$~\cite{chruscinski2006partially}.
Therefore, the assemblage defined by
\begin{equation}
    \tau_{a|x} := \Tr_A((M_{a|x}\otimes \id)\rho_\Lambda)^T
\end{equation}
is $n$-preparable~\cite{designolle2021genuine}.
Hence,
\begin{align}
    \sum_{a,x} \Tr(\Lambda(\rho_{a|x})M_{a|x}) 
     = \sum_{a,x} \Tr(\rho_{a|x}\Lambda^*(M_{a|x})) 
    = d \sum_{a,x} \Tr(\rho_{a|x} \tau_{a|x}) \: .
\end{align}
Choosing $\rho_{a|x}$ to be the known HD steering witness $\rho_{a|x} = \ket{e_{a|x}}\bra{e_{a|x}}$ where $\{\ket{e_{a|x}}\}_a$ are MUBs~\cite{designolle2021genuine}, implies that
\begin{align}
    \sum_{a,x} \Tr(\Lambda(\rho_{a|x})M_{a|x})
    = d \sum_{a,x} \Tr(\rho_{a|x} \tau_{a|x}) 
     \leq \left(\frac{\sqrt{n}-1}{\sqrt{n}+1}+1 \right) \left(\frac{1}{\sqrt{d}}+1 \right) d \: .
\end{align}
This witness requires partial trust which means that the measurements $\{M_{a|x}\}_a$ of the remote party are not assumed to have a quantum description.
Using the definition of $C_{ab}^x$ we recover the channel PT witness from Eq.~\ref{eq:channel schmidt number bound}, i.e.
\begin{equation}
    \sum_{a,x} C_{a,a|x}  \leq \frac{2 \sqrt{n} (d + \sqrt{d})}{\sqrt{n} + 1}.
\end{equation}

\section{Experimental Details} \label{SM:experiment}

\subsection{Details of the Setup} \label{SM:exp_details}
The experiment consists of three stages: state generation, propagation through the multi-mode fiber channel, and measurement. The apparatus used for the state generation consists of a spectrally filtered laser source at $810 \pm 1.5$~nm launched into a single mode fiber (Thorlabs-780HP), followed by a set of lenses with effective focal length $L1 = 59$~mm used for collimating the light, a Spatial Light Modulator (SLM) (Hamamatsu LCOS-X10468) and finally another set of lenses with effective focal length $L2=22$~mm. The light is coupled into a GRIN MMF and transmitted through the fiber channel. We test two optical fibers with different lengths: a 2~m-long fiber (Thorlabs M116L02; core diameter, $50.0 \pm2.5 \mu$m; numerical
aperture, $0.200 \pm 0.015$) and a 5~m-long fiber (Thorlabs M116L05; core diameter, $50.0 \pm2.5 \mu$m; numerical aperture, $0.200 \pm 0.015$). At the output of the fiber, we perform the measurement. The measurement apparatus consists of another lens system (with effective focal length $L3=22$~mm), a second SLM, a final lens system (with effective focal length $L4 =33$~mm) and a CMOS camera (XIMEA-xiC USB3.1). 

\subsection{Experimental Procedure} \label{SM:exp_procedure}

The state generation is performed by Alice. It consists of a Gaussian laser beam reflected off of a programmable SLM. The input state is encoded in discrete macro-pixels in the transverse-spatial degree of freedom, as shown in Fig.~\ref{fig:experiment}a. We opt for this specific basis because it allows the implementation of high-quality projective measurements \cite{valencia2020high,goel2024inverse}. The statistical properties of a single-photon propagating through the set-up are identical to those obtained for a coherent state. This enables us to simplify the experiment and use a CCD camera for detection instead of a single-photon detector \cite{barnett2022single}.

The role of the SLM is to carve out a spatial mode of light using a computer-generated hologram. Alice generates sets of modes, $\left\{\ket{e^{\text{in}}_{b|x}}\right\}_{b,x}$, in a certain basis, $x$, corresponding to the states $\rho_{b|x} = \ketbra{e^{\text{in}}_{b|x}}{e^{\text{in}}_{b|x}}$ and sends them individually through the MMF. After propagation through the MMF, the output modes are sent to Bob. Bob’s measurement comprises another SLM and a CCD camera in its focal plane. The function of Bob's device is to project the incident light onto a set of measurement modes $\left\{\ket{e^{\text{out}}_{a|x}}\right\}_{a,x}$. The SLM here does the exact time-reversed function of Alice's SLM. By displaying the conjugate of the measurement mode $\ket{e^{\text{out}}_{a|x}}$ on the SLM, we can convert a given incident mode 
into a Gaussian mode directed to the first-order diffraction spot on the camera, the intensity of which records the measurement, $M_{a|x} = \ketbra{e^{\text{out}}_{a|x}}{e^{\text{out}}_{a|x}}$.

The relevant input-output correlations for the FT and PT witnesses are those between Alice's prepared states and Bob's measurements in two MUBs. However, due to the complex scattering effects of the MMF, finding the appropriate bases is non-trivial. Here, we use the SVD-basis of the fiber and its second mutually unbiased basis. To find the SVD-basis, we approximate the spatial-optical transformation of the MMF, i.e. with a single transmission matrix ($T_{\text{approx}}$)~\cite{mounaix2017temporal,carpenter2013110x110}. We know that our channel may be impure and therefore contain more than one Kraus operator, however, we construct the approximate transmission $T_{\text{approx}}$ to find a good estimate of an SVD-basis. To characterise $T_{\text{approx}}$, we make use of a multi-plane neural network (MPNN) \cite{goel2023referenceless}, training with the dataset obtained from state preparation and measurement on random bases $|x\rangle$ and $|y\rangle$. To achieve this, uniformly distributed random phases are displayed on each SLM at input and output of the fiber and the corresponding intensities are measured by the CMOS camera. The generated dataset 
\begin{equation}
    C = \tr_{\lambda}\bigl[\ketbra{y} \mathcal{T}_{true}\bigl(\ketbra{x}\bigr)\bigr]
     \approx |\langle y|T_{approx}|x\rangle |^{2}
\end{equation}
is constructed, where $\mathcal{T}_{\text{true}}$ is the true, impure channel (capturing the modal dispersion and mixing due to the full multispectral response of the fiber) and $\mathcal{T}_{\text{approx}}$ is a Kraus-rank one channel described by a single transmission matrix (Kraus operator), that we approximate using the MPNN.

We highlight here that the characterisation of $T_{\text{approx}}$ is used to find the SVD-basis of the channel, allowing us to optimise the violation of the dimensionality witnesses. This is similar to \cite{gutierrez2023tailoring}, where they characterise the TM for increased knowledge of aberration and misalignment effects. Here, one could avoid the characterisation of $T_{\text{approx}}$ entirely by making an intelligent choice of bases. A possible candidate would be propagation-invariant modes, as used in  \cite{butaite2022build} and characterised in \cite{ploschner2015seeing}. Another potentially more promising candidate is principal modes, these are a wavelength-independent mode basis to the first order derivative that can be thought of as eigenmodes of the group-delay operator~\cite{carpenter2017comparison,fan2005principal}.

Once $T_{\text{approx}}$ is characterised, we perform a singular value decomposition $T_{\text{approx}}= UDV^{\dagger}$ to obtain two unitary matrices $U$ and $V$ which tend to diagonalise the channel and from which we can construct the approximated eigenbases. We denote the $b^{\text{th}}$ input state in this basis ($x=0$) as $|e_{b|0}^{\text{in}}\rangle = \sum_{i} V_{ib} |i\rangle$ and the $a^{\text{th}}$ output state in this basis as $|e_{a|0}^{\text{out}}\rangle = \sum_{j} U_{ja} |j\rangle$, where $V_{ib}$ is the $ib$-entry of the unitary $V$, $U_{ja}$ is the $ja$-entry of the unitary $U$ and $|i\rangle$ and $|j\rangle$ are elements of the standard basis. Our choice of standard basis in this experiment is the macro-pixel basis. If the fiber was fully described by $T_{\text{approx}}$ one would observe perfect correlations when preparing an input state $|e_{b|0}^{\text{in}}\rangle$ and measuring an output $|e_{a|0}^{\text{out}}\rangle$:
\begin{align*}
    C_{a,b|0} &= \left\|\langle e_{a|0}^{\text{out}}| T_{\text{approx}} |e_{b|0}^{\text{in}}\rangle \right\|^{2} \\
    &= \left\|\sum_{i,j} U_{ja}^{*}  V_{ib} \langle j|UDV^{\dagger}|i\rangle \right\|^{2} = \left\|D_{ab} \right\|^{2},
\end{align*}
where we use that $T_{\text{approx}}=UDV^{\dagger}$. Although we know that the channel is impure and therefore not fully described by $T_{\text{approx}}$ (we explore this in more detail in Appendix~\ref{SM:sims}), we choose these correlations to form the first measurement for the two witnesses. 

To perform the second measurement, we find bases that are mutually unbiased to $\{|e_{b|0}^{\text{in}}\rangle\}_{b}$ and $\{|e_{a|0}^{\text{out}}\rangle\}_{a}$, respectively. Given a unitary matrix, $W^{(x)}$, defining a basis mutually unbiased to the standard basis, we transform the eigenbases so that the $b^{\text{th}}$ input state in this $x$-th MUB is $\ket{e_{b|x}^{\text{in}}}= \sum_{l}W^{(x)}_{lb}|e_{l|0}^{\text{in}}\rangle$ and the $a^{\text{th}}$ output state is $\ket{e_{a|x}^{\text{out}}}= \sum_{k}W^{(x)}_{ka}|e_{k|0}^{\text{out}}\rangle$, where $W^{(x)}_{lb}$ ($W^{(x)}_{ka}$) is the $lb$-entry ($ka$-entry) of the matrix $W^{(x)}$, and we may consider $W^{(0)}=\mathbb{1}$. The correlations (if $T_{\text{approx}}$ was a perfect description of the channel) when measuring in this basis are:
\begin{align*}
    C_{a,b|x} &= \left\|\langle e_{a|x}^{\text{out}}| T_{\text{approx}}|e_{b|x}^{\text{in}} \rangle \right\|^{2} \\
    &= \left\| \sum_{l,k,i,j} W_{ka}^{(x)*} W^{(x)}_{lb} U_{ja}^{*}V_{ib} \langle j| UDV^{\dagger} |i\rangle \right\|^{2}\\
    &= \left\|(W^{(x)\dagger} D W^{(x)})_{ab} \right\|^{2},
\end{align*}
with the normalisation $\sum_a C_{a,b|x}=1$ for each $j$ and $x$. For idealised (lossless, pure and thus unitary) fiber channel, $D=\mathbb{1} \Leftrightarrow C_{x}=\mathbb{1}$. 
These correlations allow the evaluation of both the FT and PT witnesses of the channel dimensionality.

Maintaining coherence between large sets of modes is challenging, particularly given that higher order modes of fibers are in practise more lossy and dispersed. Therefore, investigating subspaces of a channel, restricted to some subset of lower order modes, can inform us how best to coherently transmit states through the channel. We explore this idea by repeating the procedure for channel subspaces of different dimension $d$. For each subspace dimension, we retain only the leading elements of the SVD-basis $\{|e_{b|0}^{\text{in}} \rangle \}_{b=1,...,d}$, and construct the MUBs using matrices $W^{(x)}$ of the corresponding dimension. The results of which are displayed in the main text in Fig.~\ref{fig:results}.

\subsection{Comparison of FT witnesses}

As discussed in the main text and Appendix~\ref{SM:FTwitness_derivation}, any entanglement certification witness can be translated into the channel picture. As such, we presented two FT witnesses (Eq.~\ref{eq:Mubbywitness} and Eq.~\ref{eq:FTwitness}). In Fig.~\ref{fig:mubbyvsmorelli}, we compare these two witnesses using the experimental data. In both the 2~m (a) and 5~m (b) cases, there is a deviation between the two witnesses for larger subspace dimensions, we found that the Baveresco witness certifies a larger dimensionality. In the 5~m case, after $d=150$, both witnesses fail to certify any dimensionality as the dispersion in the fiber is too great. 

\begin{figure*}[h!]
    \centering
\includegraphics[width=\textwidth]{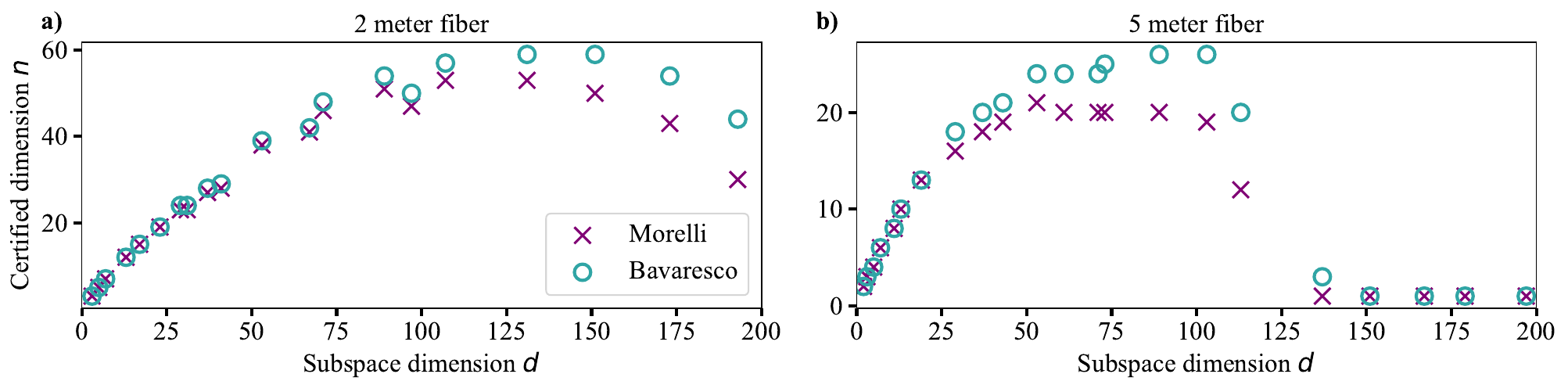}
\caption{Experimentally certified dimension $n$ of 2~m and 5~m graded-index multi-mode fibers, using the FT Bavaresco witness, Eq.~\ref{eq:Mubbywitness} (blue circles) and the FT Morelli witness, Eq.~\ref{eq:FTwitness} (purple crosses) when utilising subspaces of different dimensions $d$. 
}\label{fig:mubbyvsmorelli}
\end{figure*}

\section{Simulation Details} \label{SM:sims}

In this appendix, we explain how we simulate the entire experimental process for the scenario in which the quantum channel is an idealised straight MMF (only limited by dispersion).

\subsection{Constructing an MSTM for an idealised fiber} \label{SM:MSTM}

An idealised MMF is a theoretical model of a fiber that has no curvature, rotations or imperfections. Of course, no fabricated fiber is completely free of imperfections, and it is not possible to ensure that a fiber of the lengths we consider here is perfectly straight without any bends or rotation. 
Nevertheless, we are motivated to study this case in order to establish an upper bound on the channel dimensionality that one might expect from such idealised fibers.

The eigenmodes of the idealised fiber do not couple at all. Therefore, the transmission matrix of an idealised fiber, in the eigenmode basis, is diagonal – i.e. there is no spatial-coupling between modes, and only modal dispersion appears. For such a fiber, and choosing $\ket{e^{\text{in}}_{b|0}}$ and $\ket{e^{\text{out}}_{a|0}}$ from the SVD-basis of the TM, the correlations, $C_{a,b|0}$, Eq.~\ref{eq:statistics} are perfectly diagonal. Although the eigenmodes do not couple during fiber propagation, each eigenmode gains a relative phase difference continuously changed across the wavelength $\lambda$, corresponding to its temporal delay. For a graded-index MMF, the eigenmodes are grouped into mode groups – all modes within a mode group propagate with the same group velocity and hence acquire the same phase when exiting the fiber. 

To model these spectral effects, one can construct a stack of $\{T(\lambda_{n})\}_{n}$ transmission matrices, constituting a Multi-Spectral Transmission Matrix (MSTM)~\cite{carpenter2015observation,mounaix2016spatiotemporal}. This construction discretises the spectral dependencies, ensuring that the difference $\delta\lambda = \lambda_{n+1} - \lambda_n$ remains smaller than the spectral bandwidth of the idealised MMF. This means the resolution of the MSTM is narrow enough to account for all the spectral dependencies and the integral $\int d\lambda f(\lambda)$ can be approximated by the sum $\tfrac{1}{N}\sum_n f(\lambda_n)$, where $f(\lambda)$ is the spectrum of a light source.

The first stage of the simulation is constructing an MSTM for an ideal fiber, with a bandwidth that corresponds to the input light in the experiment. For each considered wavelength $\lambda$, a monochromatic transmission matrix describes the spatial mode coupling. In the eigenmode basis, an ideal fiber has a diagonal TM, where the phases along the diagonal correspond to the acquired phase of each mode \cite{shemirani2009principal, snyder1983optical}:
\begin{equation} 
T(\lambda)=\left[\begin{array}{ccc}
e^{-i \beta_{00}^\lambda L} & & 0 \\
& \ddots & \\
0 & & e^{-i \beta_{m n}^\lambda L}
\end{array}\right]
\end{equation}
where $L$ is the fiber length and $\beta_{mn}^{\lambda}$ is the propagation constant at wavelength $\lambda$, for a given mode, described by mode indices $m$, $n$. The propagation constants are given by:
\begin{equation}
    \beta_{m n}^{\lambda}=\frac{1}{r} \sqrt{\left(n_{1} k r\right)^2-\tilde{B}},
\end{equation}
where
\begin{equation}
\tilde{B}=\left(\frac{\Gamma\left(\frac{1}{\alpha}+\frac{1}{2}\right)(\alpha+2)(m+n+1) \pi^{1 / 2} V^{\frac{2}{\alpha}}}{2 \Gamma\left(\frac{1}{\alpha}\right)}\right)^{\frac{\alpha}{\alpha+2}}.
\end{equation} 
$\alpha$ is the power law exponent, taken here to be 2 for parabolic index core, $\Gamma$ is the gamma function and $V= k r n_{1} \sqrt{(2\Delta)}$, where $k=\frac{2\pi}{\lambda}$ is the wavenumber, $r$ is the core radius, $n_{1}$ is the refractive index of the fiber core and $\Delta = \left(n_1^2-n_{2}^2\right) /\left(2 n_1^2\right)$ is the refractive index contrast where $n_{2}$ is the refractive index of the fiber cladding. 

In our simulation, we construct an MSTM using the definitions above. The constructed MSTM is a stack of 201 TMs, with a bandwidth of $3$~nm, centered at $\lambda_{0} = 810$~nm, such that $\lambda_{n} \in [808.5, 811.5]$~nm and $\delta \lambda =0.015$~nm. In Table~\ref{tab:MSTM_parameters} below, we define all parameters used for the construction of the idealised fiber MSTM.

\begin{table}[H]
    \begin{center}
    \begin{tabular}{ |c|c| } 
     \hline
     Parameter & Simulation Value  \\ 
     \hline
    Length of fiber ($L$) & 2~m and 5~m \\
    Core radius ($r$) & 25~$\mu$m  \\ 
    Refractive index of core ($n_{1}$) & 1.444 \\ 
    Numerical aperture (NA) & 0.22 \\ 
    Central wavelength ($\lambda_{0}$) & 810~nm \\ 
    Spectral bandwidth ($\delta \lambda)$ & 0.015~nm \\ 
     \hline
    \end{tabular}
    \caption{Simulation parameters.}
    \end{center}
    \label{tab:MSTM_parameters}
\end{table}

\subsection{Reconstructing $T_{\text{sim}}$} \label{SM:TM_sim}

In the experiment, we measure a single transmission matrix $T_{\text{approx}}$ to find bases that diagonalise the fiber (see Appendix~\ref{SM:exp_procedure}). This process, is, in effect, averaging over the spectral degree of freedom and hence may not extract the optimal bases. This is because the quantum channel is impure and therefore a single transmission matrix does not capture the full spatio-spectral coupling of the fiber. To understand the effect of ideal modal dispersion on the measurement statistics, we need to simulate the averaging process performed in the experiment. 

Following the construction of the ideal fiber MSTM, we simulate the reconstruction of the approximated transmission matrix $T_{\text{sim}}$. A set of random measurements are simulated, denoted $x$ and $y$, generating a dataset $C(\lambda) = |\langle y|T(\lambda)|x\rangle |^{2}$ for each $\lambda$ in the MSTM stack. An average over $\lambda$ is taken, to give a single dataset 
$$C^{\text{avg}} = \sum_{\lambda} |\phi_{\lambda}|^{2} |\langle y|T(\lambda)|x\rangle |^{2},$$
where we weight each $T(\lambda)$ in the MSTM according to a Gaussian spectral profile across the bandwidth of the laser $|\phi_{\lambda}|^{2} =e^{\left(-\lambda^2 / 2\sigma^2\right)}/\sigma\sqrt{2 \pi}$. This corresponds to the dataset recorded in the experiment. We characterise the simulated TM, $T_{\text{sim}}$ by plugging $C^{\text{avg}}$ into the MPNN in the same way as in the experiment \cite{goel2023referenceless}.

Once we have constructed $T_{\text{sim}}$, we perform a singular value decomposition (SVD) $T_{\text{sim}}= UDV^{\dagger}$. We simulate Bob's measurement in the $x^{\text{th}}$ MUB, generating a correlation matrix for each TM in the MSTM, given by 
\begin{align}
    C_{a,b|x}^{(\lambda)} &= \left\|\langle e_{a|x}^{\text{out}}| T(\lambda)|e_{b|x}^{\text{in}} \rangle \right\|^{2} \\
    &= \left\| \sum_{l,k,i,j} W^{(x)*}_{ka} W^{(x)}_{lb} U_{ja}^{*}V_{ib} \langle j| T(\lambda) |i\rangle \right\|^{2}
\end{align}
where $U$ and $V$ are from the SVD of $T_{\text{sim}}$. Again, we average over the spectral degree of freedom, weighted by the Gaussian profile of the laser, giving an average correlation matrix:
\begin{align} \label{eq:C_avg}
    C^{\text{avg}}_{a,b|x} = \sum_{\lambda} |\phi_{\lambda}|^{2} C_{a,b|x}^{(\lambda)}.
\end{align}
This is normalised and then processed using the FT and PT witnesses. 

\subsection{Noisy Channel Simulation} \label{SM:sim_noisy}

In this appendix, we introduce a model to characterise the noise levels in our experiment. The ideal channel in Kraus representation is: 
\begin{equation}
    \Lambda_{\text{ideal}}(\rho_{in}) = \sum_{\lambda} T_{\lambda} \rho_{in} T_{\lambda}^{\dagger},
\end{equation}
where $\sum_{\lambda} T_{\lambda}^{\dagger} T_{\lambda} = \mathbb{1}$, $T_{\lambda} = |\phi_{\lambda}|T(\lambda)$ and $\sum_{\lambda}|\phi_{\lambda}|^{2} = 1$. We add noise to the channel by mixing the ideal channel with a channel that takes every state to the maximally mixed state $\Lambda_{\text{mm}}(\rho) = \frac{\mathbb{1}}{d}$, with some mixing parameter $p$:
\begin{equation}
    \Lambda_{\text{noisy}} = p \Lambda_{\text{ideal}} + (1-p) \Lambda_{\text{mm}}.
\end{equation}
where $0 \leq p \leq 1$. Equivalently, we can write the average correlation matrix that would result from the channel $\Lambda_{\text{noisy}}$ in terms of the correlations from the two channels (idealised and noisy):
\begin{equation}
    C_{\text{noisy}} = p C_{\text{ideal}} + (1-p)C_{\text{mm}},
\end{equation}
where $C_{\text{ideal}}$ is given by Eq.~\ref{eq:C_avg} and $C_{\text{mm}}$ is a $d\times d$ matrix whose elements all equal $\frac{1}{d}$. 

Importantly, as the considered subspace dimension increases, the noise in the experiment increases and this is due to experimental factors such as
mode-dependent loss, misalignment, errors in the state preparation, measurements and the transmission matrix estimation. Therefore, we consider a noise model in which the mixing parameter $p$ increases with respect to subspace dimension. We fit a quadratic variation of $p$ with respect to subspace dimension $d$:
\begin{align}
    p = ad^2 + bd  + c,
\end{align}
where the coefficients $a$, $b$ and $c$ are summarised in Table~\ref{tab:quad} for the 2~m and 5~m fiber cases. We note that this noise model is designed for our experimental dataset and is not physically motivated, hence we do not expect it to hold outside of the dimension ranges considered here.
\begin{table}[H]
    \begin{center}
    \begin{tabular}{ |c|c|c| } 
     \hline
     Coefficients & 2~m fiber & 5~m fiber \\ 
     \hline
    $a$ & $7.415 \times 10^{-6}$ & $6.167\times 10^{-6}$ \\
    $b$ & $-2.851 \times 10^{-3}$ & $-2.549\times 10^{-3}$  \\ 
    $c$ & $9.864\times 10^{-1}$ & $8.769\times 10^{-1}$ \\ 
     \hline
    \end{tabular}
    \caption{Polynomial coefficients for the noisy simulation.}
    \end{center}
    \label{tab:quad}
\end{table}

\section{Example scenario}

Finally, in Fig.~\ref{fig:media_example}, we present a fictional scenario in which the PT witness could be used. A video streaming company wishes to distribute content to multiple users in high-definition. Our PT witness provides a quick test to verify that the fibers connecting their trusted system to the untrusted end-users can transmit the data in the dimension required, without having to trust that the output measurement devices are functioning perfectly.

\begin{figure*}[h!]
    \centering
\includegraphics[width=0.7\textwidth]{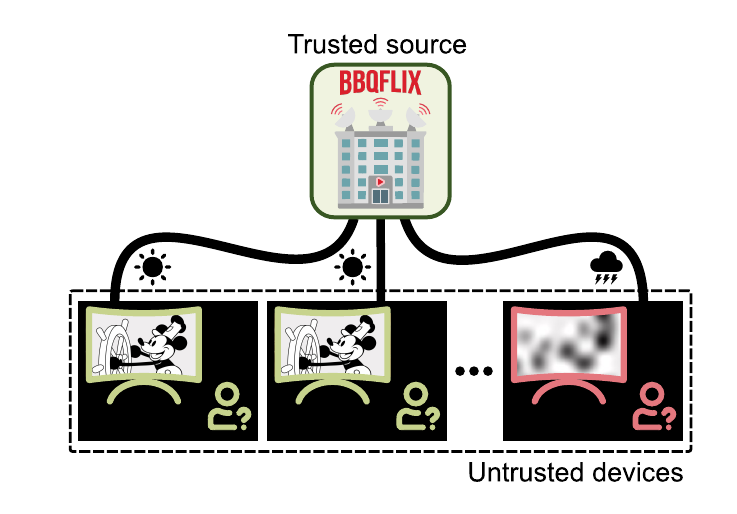}
\caption{Example scenario 
}\label{fig:media_example}
\end{figure*}

\end{widetext}

\end{document}